\documentclass[aps,twocolumn,prd,showpacs,nofootinbib]{revtex4}
\usepackage{amsmath}
\usepackage{graphicx}
\usepackage{dcolumn}
\usepackage{bm}
\usepackage{amssymb}
\usepackage{latexsym}

\def\setC{\mathbb{C}}

\def\setR{\mathbb{R}}

\newcommand{\dd}{\mathrm{d}}
\newcommand{\ee}{\mathrm{e}}

\newcommand{\ie}{\textsl{i.e.~}}

\newcommand{\eg}{\textsl{e.g.~}}

\newcommand{\GReCO}{${\cal G}\setR\varepsilon\setC{\cal O}$}

\def\spose#1{\hbox to 0pt{#1\hss}}
\def\lta{\mathrel{\spose{\lower 3pt\hbox{$\mathchar"218$}}
     \raise 2.0pt\hbox{$\mathchar"13C$}}}
\def\gta{\mathrel{\spose{\lower 3pt\hbox{$\mathchar"218$}}
     \raise 2.0pt\hbox{$\mathchar"13E$}}}

\def\O{g_{_\mathrm{II}}}
\def\E{f_{_\mathrm{II}}}
\def\AI{A_{_\mathrm{I}}}
\def\BI{B_{_\mathrm{I}}}
\def\AIII{A_{_\mathrm{III}}}
\def\BIII{B_{_\mathrm{III}}}
\def\TB{T_{_\mathrm{B}}}
\def\RB{R_{_\mathrm{B}}}

\begin{document}

\title{On the properties of the transition matrix in bouncing
cosmologies}

\author{J\'er\^ome Martin}
\email{jmartin@iap.fr}
\affiliation{Institut d'Astrophysique de
Paris, \GReCO, FRE 2435-CNRS, 98bis boulevard Arago, 75014 Paris,
France}

\author{Patrick Peter}
\email{peter@iap.fr}
\affiliation{Institut d'Astrophysique de
Paris, \GReCO, FRE 2435-CNRS, 98bis boulevard Arago, 75014 Paris,
France}

\date{January 27$^\mathrm{th}$, 2004}

\begin{abstract}
We elaborate further on the evolution properties of cosmological
fluctuations through a bounce. We show this evolution to be
describable either by ``transmission'' and ``reflection'' coefficients
or by an effective unitary $S-$matrix. We also show that they behave
in a time reversal invariant way. Therefore, earlier results are now
interpreted in a different perspective and put on a firmer basis.
\end{abstract}

\maketitle

\section{Introduction}

A (Jordan or Einstein frame-) bounce~\cite{bounce} is predicted to
have taken place in some string -- or quantum gravity -- motivated
early universe models, as \eg the pre big bang~\cite{PBB} and
ekpyrotic cases~\cite{ekp,noekp}. Those models are often presented as
plausible challengers to the otherwise well-confirmed inflation
paradigm, and must therefore be able to provide an almost
scale-invariant perturbation spectrum.

The difficulty resides in the propagation of the spectrum, usually
produced in the contracting phase, through the bounce, as the latter
often involves high curvatures and thus knowledge of at least some
non-linear terms in the underlying theory.  Even assuming that the
perturbations can in fact somehow propagate through, \ie supposing the
first order to remain meaningful all along, a point quite debatable in
itself (see \eg Lyth in Refs.~\cite{noekp}), it is necessary to impose
a specific model to examine the validity of the usual assumptions that
are made, either explicitly or implicitly, in the abovementionned
frameworks.

This has been done in Refs.~\cite{MPD,MPL} in which we used pure
general relativity, a Friedmann-Robertson-Walker metric with
positively curved spatial sections and a scalar field $\varphi$ to
provide the matter content and lead the dynamics of the scale factor
$a(\eta)$ as a function of conformal time $\eta$. We have since
received inquiries about some seemingly unexpected features of the
results, such as an apparently vanishing-determinant transition
matrix, and we were questioned about the unitarity of the evolution.

In this brief report we clarify these issues by showing that, since
the evolution equation can be understood as a time-independent
Schr\"odinger equation, its solutions can be interpreted as
transmission and reflection coefficients, which ensures the
conservation of probability, or, stated differentely, the unitarity of
the $S-$matrix that can be built out of the transition. Note however
that these results stem from the mathematical properties of the
Schr\"odinger equation, although the actual interpretation of the
underlying classical perturbed Einstein equations does involve neither
any physical wavefunction nor any probabilistic meaning. In
particular, this should not be confused with the effective
quantization of the physical modes, \ie the second quantization of the
semi-classical effective field theory which underpins inflationary
structure formation scenarios.

In addition, we show that the transition matrix reflects the
underlying invariance under time reversal. Finally, we also
demonstrate that the vanishing of the determinant signaled before was
to be expected and turns out in fact to provide a necessary
consistency check of the calculations performed in Ref.~\cite{MPD}.

\begin{figure*}[th]
\includegraphics[width=7cm]{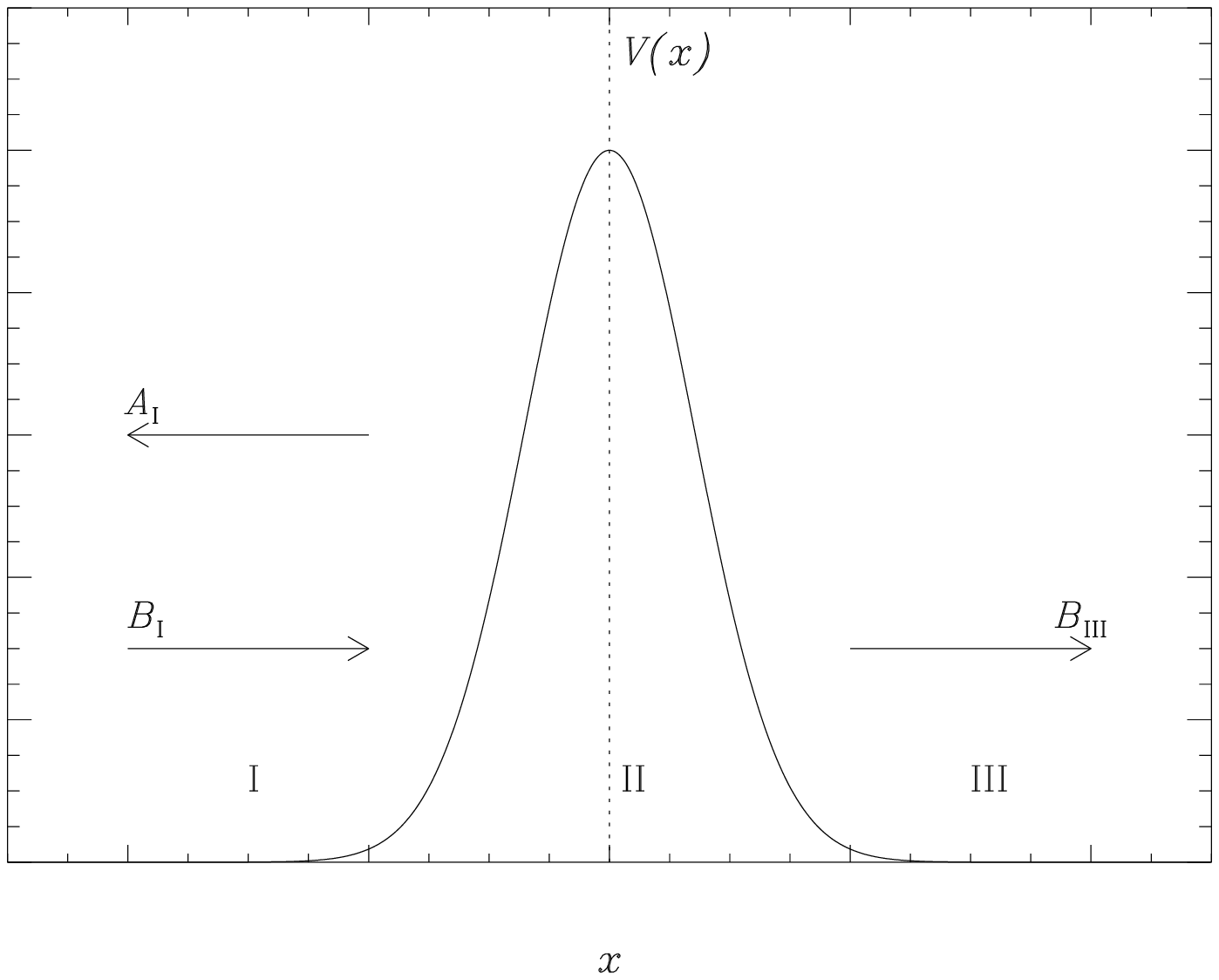}\hskip1cm
\includegraphics[width=7cm]{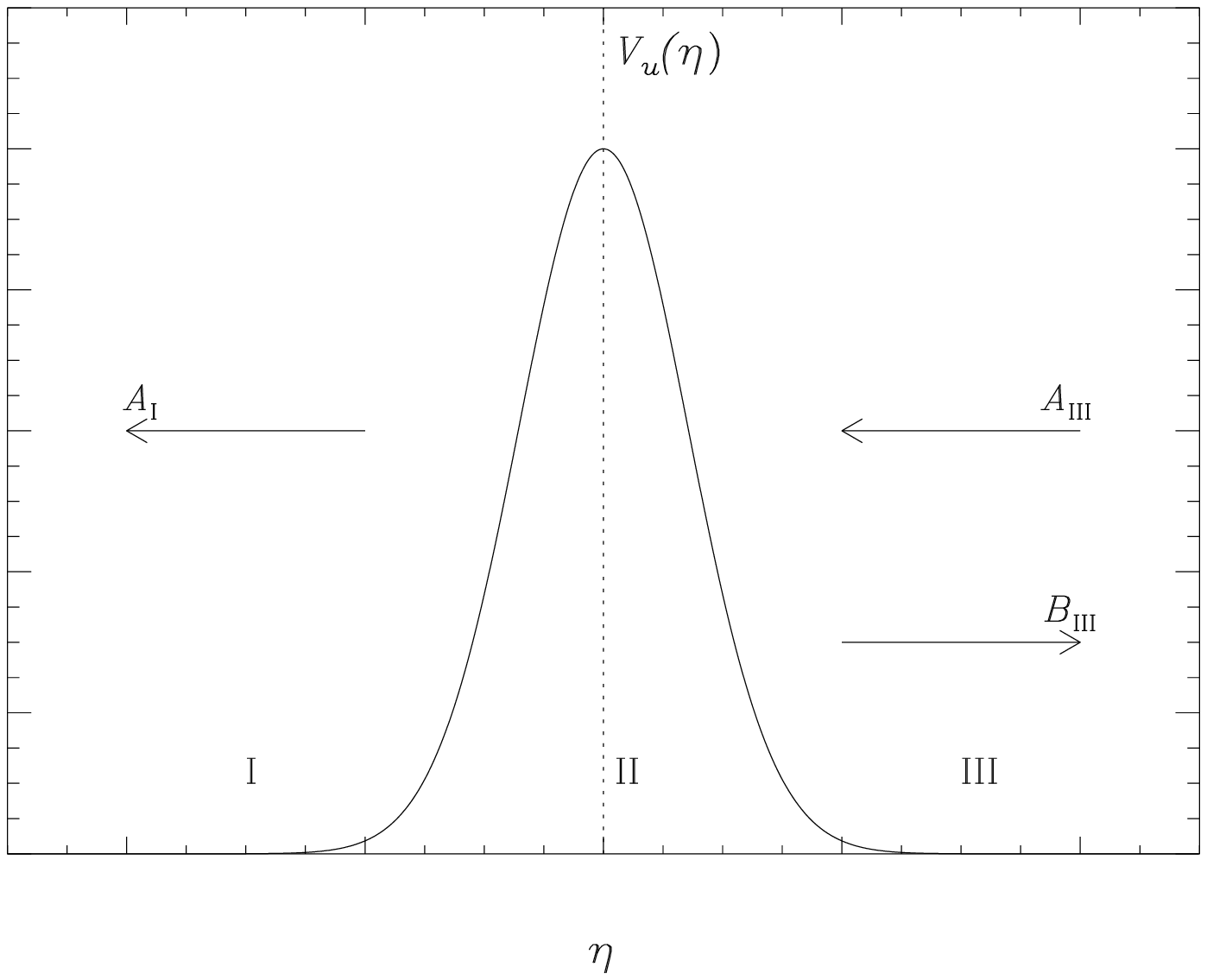}
\caption{The analogy and differences between the wavefunction across a
potential in ordinary quantum mechanics and primordial perturbations
through a bounce. In both cases, the total current is conserved.}
\label{fig:potentials}
\end{figure*}

\section{The bounce seen as a potential of a Schr\"odinger equation}

We want to describe the evolution of cosmological perturbations in a
universe with positively curved spatial sections through a bounce
described by general relativity and a scalar field $\phi$. The scale
factor $a$ depends on the conformal time $\eta$.  One can construct a
perturbation function $u\propto \Phi$, proportional to the Bardeen
potential $\Phi$ through a function of the background~\cite{MPD} which
we do not need to specify here, and satisfying~\cite{MPD,MPL}
\begin{equation}\label{Schroe}
u'' + \left[ k^2 - V_u(\eta)\right] u = 0,
\end{equation}
for a wavenumber $k=\sqrt{n(n+2)}$ on the 3-sphere. By setting $k^2\to
E$, $u \to \psi$ and $\eta\to \sqrt{2m} x/\hbar$, Eq.~(\ref{Schroe})
is of the form of a time-independent Schr\"odinger equation
$-(\hbar^2/2m) \dd^2\psi/\dd x^2 + \left[V(x)-E\right]\psi=0$ in the
variable $\eta$ for the ``wavefunction'' $u$ in the potential
$V_u(x)$. Given this form, the current $\mathcal{J} = -(i\hbar/2m)
\left (\psi^\star\nabla \psi - \psi\nabla\psi^\star\right) \to  -
i\left(u^\star u'- u u^{\star\prime}\right)$ is conserved
$\nabla\cdot\mathcal{J}= 0\to\mathcal{J}'=0$.

In the case of a symmetric bounce, the potential $V_u(\eta)$ of
Eq.~(\ref{Schroe}) is nonvanishing in a finite and symmetric region
around $\eta =0$, \ie for $-\eta_\mathrm{z}\leq \eta\leq
x_\mathrm{z}$, which we shall call region II, surrounded by regions I
(for $\eta$ before the bounce) and III (for $\eta$ after the
bounce). In both regions I and III, the potential being negligible, we
have
\begin{equation}
u_{_\mathrm{I,III}} = \frac{1}{\sqrt{2k}} \left( A_{_\mathrm{I,III}}
\ee^{-ik\eta} + B_{_\mathrm{I,III}} \ee^{ik\eta}\right),
\end{equation}
leading to the currents
\begin{equation}
J_{_\mathrm{I,III}} =\left| B_{_\mathrm{I,III}}
\right|^2 -\left| A_{_\mathrm{I,III}}\right|^2 
\end{equation}
on both sides of the bounce. Apart from the overall irrelevant
normalization, this is understood, in ordinary quantum mechanics, as
the difference between current probabilities of wavefunctions
traveling respectively in the $-\eta$ (coefficient $A$) and $+\eta$
(coefficient $B$) directions. The transition matrix relating
$(\AIII,\BIII)$ to $(\AI,\BI)$ was obtained in Refs.~\cite{MPD,MPL} by
expanding the solution in the region II into an even $\E (-\eta)=\E
(\eta)$ and an odd $\O (-\eta)=-\O (-\eta)$ functions, which can
always be done in the case of a symmetric potential.

Setting $\E\equiv \E (\eta_\mathrm{z})$, $\E'\equiv \dd \E
(\eta_\mathrm{z})/\dd \eta$, $\O\equiv \O (\eta_\mathrm{z})$, and
$\O'\equiv\dd\O(\eta_\mathrm{z})/\dd \eta$, we found
\begin{equation}
\begin{pmatrix} 
\AIII \cr \BIII
\end{pmatrix} 
= \begin{pmatrix} X & Y \cr Y^\star & X^\star
\end{pmatrix}
\begin{pmatrix}
\AI \cr \BI
\end{pmatrix} \equiv T_u \begin{pmatrix}
\AI \cr \BI
\end{pmatrix} \, ,
\label{transT}
\end{equation}
thereby defining the transition matrix $T_u$, where
\begin{eqnarray}
X&=&\frac{i\ee^{2ikx_\mathrm{z}}}{k} \frac{\left(
\E'-ik\E\right)\left(\O'-ik\O\right)}{\left( \E\O'-\O\E'\right)},\\
Y&=&\frac{i}{k}\frac{\left(\E'\O'+k^2\O\E\right)}{\left(\E\O'-\O\E'\right)},
\end{eqnarray}
both terms being calculated at the points of matching.

{}From Eq.~(\ref{transT}), one sees that the determinant of $T_u$ is
unity, namely $|X|^2 - |Y|^2 =1$, as may be checked by direct
calculation. This property is related to the fact that the propagation
of perturbations through a bounce is similar to that of a wavefunction
$\psi(x)$ through a potential $V(x)$, as illustrated on the right of
Fig.~\ref{fig:potentials}. If the incoming wavefunction has amplitude
$\BI$, it is reflected with amplitude $\AI$, leading to a reflection
coefficient $R=|\AI/\BI|^2$, and the amount $\BIII$ goes through, so
that the transmission coefficient is $T=|\BIII/\BI|^2$ (assuming the
same wavenumber on both sides). For the bounce case shown on the left,
the analogy makes sense provided one reads the figure backwards in
time. Then, the ``transmitted'' wave, with amplitude $\AI$, in fact
the original perturbation, travels again in the same direction as the
wave hitting the potential with amplitude $\BIII$, leading to a
transmission coefficient $\TB\equiv |\AI/\AIII|^2$, while the
reflection coefficient reads $\RB\equiv |\BIII/\AIII|^2$. Then, the
current conservation, $\mathcal{J}'=0$, implies that
\begin{equation}
\RB + \TB =1,
\end{equation}
as expected. Notice that this is just a rewriting of the matching
condition $|\BIII|^2-|\AIII|^2 = |\BI|^2-|\AI|^2$, itself a
consequence of the fact that the transition matrix is of unit
determinant: this merely expresses the conservation of probability in
ordinary quantum mechanics. The result obtained in Ref.~\cite{MPD} is
therefore in perfect agreement with the conservation of the effective
probability current associated with the otherwise classical Einstein
equations.

\section{Bounce $S-$Matrix and Unitarity}

Another way to express conservation of probability is through
unitarity. Therefore, one must consider a different point of view,
namely that of the $S-$matrix, to build the unitary transformation
stemming from $T_u$. This is what we do in this section, by means of a
comparison with standard collision processes in ordinary quantum
mechanics. Obviously, there is no actual collision involved, and the
fact that one can construct such a unitarity $S-$matrix merely
reflects the mathematical consistency of Refs.~\cite{MPD,MPL}.

In three dimensional quantum mechanics, a plane wave propagating in an
incident direction $\mathbf{n}=\mathbf{k}/|\mathbf{k}|$ and diffusing
in a potential located at $O$ is observed at the point $M$ such that
$\vec{OM}\equiv r \mathbf{n}'$ leads to the collisional wave function
$\Psi_{\mathbf{n},\mathbf{n}'}$ given by~\cite{landau}
\begin{equation}\label{Psi}
\Psi_{\mathbf{n},\mathbf{n}'} = \ee^{ikr \mathbf{n}\cdot\mathbf{n}'} +
\frac{1}{r} 
f_{\mathbf{n},\mathbf{n}'} \ee^{ikr},
\end{equation}
thus defining the coefficients $f_{\mathbf{n},\mathbf{n}'}$
characterizing the diffusion and providing the $S-$matrix (see below),
the factor $1/r$ in the second term stemming from the expression of a
spherical wave in 3D. In a similar way, one can interpret the
transformation Eq.~(\ref{transT}) in terms of such a $S-$matrix, and
hence to check unitarity.

In the cosmological one dimensional case, the $1/r$ factor of
Eq.~(\ref{Psi}) is absent, the role of the radial variable $r$ is
played by the absolute value $|\eta|$ of the conformal time, and the
unit vectors $\mathbf{n}$ and $\mathbf{n}'$ can point towards only two
directions, namely $+\hat\eta$ and $-\hat\eta$, so that
$\mathbf{n}\cdot\mathbf{n}' = \pm 1$. In what follows, we shall denote
the time directions $\pm\hat\eta$ simply by the symbols $+$ or $-$.

Let us consider the case of the figure on the right, \ie the
cosmological case with the potential $V_u(\eta)$. In this case, the
analogous form of $\Psi$ for $u$ in Eq.~(\ref{Psi}) is
$u_{\mathbf{n},\mathbf{n}'}$. After the bounce, the solution for $u$
is
\begin{equation}
u_\mathrm{after} =\frac{1}{\sqrt{2k}}\left(\AIII \ee^{-ik|\eta|} + \BIII
\ee^{ik|\eta|}\right),
\end{equation}
from which, in analogy with Eq.~(\ref{Psi}), one gets $\mathbf{n} =
-\hat\eta$, $\mathbf{n}'=+\hat\eta$ and therefore $f_{-+}$ is simply
the factor $\BIII/\sqrt{2k}$ evaluated for $\AIII=\sqrt{2k}$ and
$\BI=0$, \ie $f_{-+}=Y^\star /X$, see Eq.~(\ref{transT}). Before the
bounce, $u$ is given by
\begin{equation}
u_\mathrm{before}=\frac{\AI}{\sqrt{2k}} \ee^{ik|\eta|} 
= \frac{1}{\sqrt{2k}}\left[\AIII \ee^{ik|\eta|} + (\AI -
\AIII) \ee^{ik|\eta|}\right] ,
\end{equation}
from which, since one still has $\mathbf{n}=-\hat\eta$ but
$\mathbf{n}'=-\hat\eta$, the comparison with Eq.~(\ref{Psi}) provides
the coefficient $f_{--}$, namely the value of $(\AI-\AIII)/\sqrt{2k}$
evaluated for $\AIII=\sqrt{2k}$, \ie $f_{--}=1/X-1$. Reasonning along
the same line but in the reverse direction for $\mathbf{n}$ gives the
coefficients $f_{++}=1/X-1$ and $f_{+-}=-Y/X$.

The total wavefunction $\Psi(M)$ is the superposition of all the
solutions for all incident directions weighted by arbitrary
coefficients $F_{\mathbf{n}}$, \ie
\begin{eqnarray}
\Psi_{\mathbf{n}'} &=& \int F_\mathbf{n} \Psi_{\mathbf{n},\mathbf{n}'} 
\dd^3 \mathbf{n} \nonumber\\ &\propto& 
\ee^{-ikr} F_{-\mathbf{n}'} - \ee^{ikr} S_{\mathbf{n},\mathbf{n}'}
F_{\mathbf{n}'}, \label{Sdef}
\end{eqnarray}
with which one defines the $S-$matrix as the coefficients in front of
the outgoing modes. Similarly, in the bouncing case, one has the
discrete sum
\begin{equation}
u_{\mathbf{n}'} =\sum_{\mathbf{n}=\pm\hat\eta} F_\mathbf{n}
u_{\mathbf{n},\mathbf{n}'},
\end{equation}
leading to
\begin{equation}
u_+ = F_-\ee^{-ik|\eta|} + \left[ f_{-+} F_- + \left( 1+f_{++}\right)
F_+\right]\ee^{ik|\eta|},
\end{equation}
and a similar formula by changing $\hat\eta\to - \hat\eta$. Gathering
all the above results, this leads a two-by-two $S-$matrix that reads
\begin{equation}
S = 1+f = \displaystyle\frac{1}{X} \begin{pmatrix}
1 & -Y\cr
Y^\star & 1
\end{pmatrix},
\end{equation}
which, because $|X|^2 - |Y|^2 =1$ as discussed in the previous
sections, is unitary ($SS^\dagger=1$). We have therefore built, from
the non-unitary transformation $T_u$, the relevant unitary $S-$matrix,
emphasizing once more the probability conservation. 

\section{Time inversion}

In Refs.~\cite{MPD,MPL}, we expanded the matrix $T_u$ of
Eq.~(\ref{transT}) in the parameter $\Upsilon$, which gives a measure
of the deviation to the null energy condition $\rho+p\geq 0$ at the
bounce. For $\Upsilon\ll 1$, the potential $V_u(\eta)$ goes to a
function akin to a Dirac $\delta$ distribution, namely $V_u(\eta) \to
- C_\Upsilon\delta(\eta)$, with $C_\Upsilon \propto \Upsilon^{-1/2}
\gg 1$. With such a potential, the perturbation $u$ crosses the bounce
unaltered, \ie $[u]=0$, while its time derivative jumps by the amount
$[u']=-C_\Upsilon u(0)$. These relations allow to obtain the matrix
exactly, and we found
\begin{eqnarray}\label{Tu}
T_u &=& \begin{pmatrix} 1-i \displaystyle\frac{C_\Upsilon}{2k} & -i
\displaystyle\frac{C_\Upsilon}{2k} \cr & \cr
i \displaystyle\frac{C_\Upsilon}{2k} & 
1+i\displaystyle\frac{C_\Upsilon}{2k} \end{pmatrix}
=1-i\frac{C_\Upsilon}{2k}
\begin{pmatrix} 1 & 1  \cr -1 & -1 \end{pmatrix}\, ,
\nonumber\\ 
\end{eqnarray}
which is indeed of the general form obtained in Eq.~(\ref{transT}). We
use this opportunity to correct a misprint in Eq.~(77) of
Ref.~\cite{MPD} in which the factor $i$ appears in the last term in
the denominator whereas it should be at the numerator.

In the presentation made in Ref.~\cite{MPD}, we explicitly assumed the
limit $n(n+2)\Upsilon\ll 1$ and thus neglected the unit matrix part of
$T_u$. Using Eq.~(\ref{Tu}), this gives
\begin{eqnarray}\label{Tuapprox}
T_u^{\rm (approx)} &=& -i\frac{C_\Upsilon}{2k}
\begin{pmatrix} 1 & 1 \cr -1 & -1 \end{pmatrix}\, ,
\end{eqnarray}
in perfect agreement with Eq.~(70) of Ref.~\cite{MPD}. As a
consequence, the determinant of the above matrix vanishes
\begin{equation}
{\rm det}\left[T_u^{\rm (approx)}\right]=0\, ,
\end{equation}
whereas, in fact, the determinant of the exact matrix is unity,
$\det\left(T_u\right)=1$.

With a time-symmetric potential $V_u(-\eta)=V(\eta)$ and only a second
order time derivative term, one expects $T_u$ to be a time reversal
invariant transformation. In other words, if the initial conditions
are no longer fixed in the region I but in the region III, the final
result, now expressed in region III and no longer in region I, should
obviously be the same, in particular should still be proportional to
$\Upsilon ^{-1/2}$. If one wishes to check this property using
$T_u^{\rm (approx)}$, one faces the problem that it is not
invertible. Hence, neglecting the term $1$ in Eq.~(\ref{Tu}) has led
to an apparent non symmetrical result while the problem clearly
possesses this symmetry. The resolution of this paradox is obvious:
the symmetry has been lost when the approximation has been performed
and $T_u$, which is invertible since $\det (T_u)\neq 0$, should be
used in this case. We are only authorized to neglect the term $1$ in
Eq.~(\ref{Tu}) if we want to obtain a close form of the transition
matrix, as done in Ref.~\cite{MPD}, but not if we want to check the
time reversal invariance of the matrix. The fact that
$\det\left[T_u^{\rm (approx)}\right]=0$ is actually nothing but a
consistency check that the calculation performed in Ref.~\cite{MPD} is
correct. Indeed, without this property, the inversion of $T_u^{\rm
(approx)}$ would have been possible, leading to an effect
$\propto\Upsilon^{1/2}$ whereas the time reversal invariance implies
that it should scale as $\Upsilon ^{-1/2}$ in both time
directions. The fact that $\det \left[T_u^{\rm (approx)}\right]=0$ can
therefore be interpreted as the ``signal sent by the theory'' that we
are not allowed to use $T_u^{\rm (approx)}$ to check the symmetry of
the system.

It remains to be checked explicitly that the matrix $T_u$ has the
correct time reversal symmetry. For this purpose, it is sufficient to
calculate the inverse of $T_u$ and one obtains
\begin{equation}
T^{-1}_u\left( C_\Upsilon \right)=T_u\left( - C_\Upsilon\right) \, .
\end{equation}
Therefore, the amplitude of the transition matrix remains proportional
to $\Upsilon ^{-1/2}$ as required. The only difference is that the
matrix of the inverse process should now be calculated with the
parameter $-C_{\Upsilon }$ rather than $+C_{\Upsilon }$. The origin of
this minus sign can easily be understood. Indeed, given the
perturbation equation (\ref{Schroe}), time inversion( in the case at
hand, given the analogy with the time-independent Schr\"odinger
equation, this operator is similar to parity) $\mathcal{T}$ manifests
itself only in the matching condition on the first time derivative
$u'$ (which is not invariant) adding an extra minus sign, namely
$\mathcal{T}([u'])=-[u]$. This is exactly equivalent to changing the
sign of $C_\Upsilon$. Comparing with Eq.~(\ref{Tu}), we see that
\begin{equation}
\mathcal{T}\left[T_u\left(C_\Upsilon\right)\right] =
T_u\left( - C_\Upsilon\right) = T^{-1}_u\left(C_\Upsilon\right),
\end{equation}
which merely expresses the familiar result that time inversion must
satisfy $\mathcal{T}^2 = \mathcal{T}$. Note that for an operator to be
an arbitrary function of some parameter such as $C_\Upsilon$, the only
way to ensure that $\mathcal{T}^2 = \mathcal{T}$ is that
$\mathcal{T}(C_\Upsilon) = -C_\Upsilon$ as indeed is the case here.

Finally, the explicit form of the $S-$matrix in the limit of
Refs.~\cite{MPD,MPL} is
\begin{equation}
S=\displaystyle\frac{1}{1-\displaystyle\frac{iC_\Upsilon}{2k}}
\begin{pmatrix}
1&\displaystyle\frac{iC_\Upsilon}{2k} \cr 
-\displaystyle\frac{iC_\Upsilon}{2k} & 1\end{pmatrix}
\simeq \begin{pmatrix} 0 & -1\cr 1 & 0\end{pmatrix} \, ,
\end{equation}
where in the last approximation we have taken advantage of the
condition $n(n+2)\Upsilon \ll 1$ as before.

\vspace{0.2cm}

\section{Conclusions} 

\vspace{-0.2cm}

We have presented a brief note on the propagation of primordial
perturbations through a bouncing phase in the early universe to
interpret this evolution in terms of either transmission and reflexion
coefficients across the potential of a non-relativistic
Schr\"odinger-like equation or in terms of the corresponding
$S-$matrix that can be constructed out of the model.  We have
demonstrated that the results obtained in Refs.~\cite{MPD,MPL} imply
that $\RB+\TB=1$ and that the $S-$matrix is unitary, thereby checking
the mathematical consistency of those results. Finally, we have also
emphasized that the structure of the evolution equation is invariant
under time reversal.

\acknowledgments 

We are grateful to R.~Brandenberger and B.~Carter for
many discussions that led to the above clarifications.

\end{document}